# Biolink Model: A Universal Schema for Knowledge Graphs in Clinical, Biomedical, and Translational Science

Deepak R. Unni*[1,2], Sierra A.T. Moxon*[2], Michael Bada[3], Matthew Brush[3], Richard Bruskiewich[4], Paul Clemons[5], Vlado Dancik[5], Michel Dumontier[6], Karamarie Fecho[7], Gustavo Glusman[8], Jennifer J. Hadlock[8], Nomi L. Harris[2], Arpita Joshi[8], Tim Putman[3], Guangrong Qin[8], Stephen A. Ramsey[9], Kent A. Shefchek[3], Harold Solbrig[10], Karthik Soman[11], Anne T. Thessen[3], Melissa A. Haendel[3], Chris Bizon[7], Christopher J. Mungall[2], and the Biomedical Data Translator Consortium

1. Genome Biology Unit, European Molecular Biology Laboratory, Heidelberg, Germany
2. Division of Environmental Genomics and Systems Biology, Lawrence Berkeley National Laboratory, Berkeley, CA, USA
3. Center for Health AI, University of Colorado Anschutz Medical Campus, Aurora, CO, USA
4. Star Informatics, Sooke, BC, Canada
5. Chemical Biology and Therapeutics Science Program, Broad Institute, Cambridge, MA, USA
6, Institute of Data Science, Maastricht University, Maastricht, The Netherlands
7. Renaissance Computing Institute, University of North Carolina at Chapel Hill, Chapel Hill, North Carolina, USA
8. Institute for Systems Biology, Seattle, WA, USA
9. Department of Biomedical Sciences, Oregon State University, Corvallis, OR, USA
10. Johns Hopkins University, Baltimore, MD, USA
11. Department of Neurology, University of California San Francisco, San Francisco, CA, USA

* Co-first authors

## Abstract

Within clinical, biomedical, and translational science, an increasing number of projects are adopting graphs for knowledge representation. Graph-based data models elucidate the interconnectedness between core biomedical concepts, enable data structures to be easily updated, and support intuitive queries, visualizations, and inference algorithms. However, knowledge discovery across these 'knowledge graphs' (KGs) has remained difficult. Data set heterogeneity and complexity; the proliferation of *ad hoc* data formats; poor compliance with guidelines on findability, accessibility, interoperability, and reusability; and, in particular, the lack of a universally-accepted, open-access model for standardization across biomedical KGs has left the task of reconciling data sources to downstream consumers. Biolink Model is an open source data model that can be used to formalize the relationships between data structures in translational science. It incorporates object-oriented classification and graph-oriented features. The core of the model is a set of hierarchical, interconnected classes (or categories) and relationships between them (or predicates), representing biomedical entities such as gene, disease, chemical, anatomical structure, and phenotype. The model provides class and edge attributes and associations that guide how entities should relate to one another. Here, we highlight the need for a standardized data model for KGs, describe Biolink Model, and compare it with other models. We demonstrate the utility of Biolink Model in various initiatives, including the Biomedical Data Translator Consortium and the Monarch Initiative, and show how it has supported easier integration and interoperability of biomedical KGs, bringing together knowledge from multiple sources and helping to realize the goals of translational science.

## Introduction

The use of graphs to formalize the representation of human knowledge dates back to the origins of Artificial Intelligence (AI) and the use of semantic networks for knowledge representation[1,2]. The term 'knowledge graph' (KG) is gaining popularity and is generally used to encompass a range of graph-oriented representation frameworks, including Resource Description Framework (RDF) triple stores and labeled property-graph databases such as Neo4j. Examples of general-domain KGs include the Google Knowledge Graph and Wikidata[3]. Within the biomedical sciences, examples include SemMedDB[4], Hetionet [5], Implicitome[6], Monarch Initiative[7], the biological subset of Wikidata[8], SPOKE[9], and KG-COVID-19[10].

While KGs have been defined in various ways, perhaps the most intuitive definition is a graph in which the nodes represent real-world entities and the edges represent known relationships between those entities[11]. In a KG, the knowledge or 'facts' are represented as statements, with each statement modeled as two nodes linked together by an edge representing the relationship between them. The statements can have additional properties, metadata, and qualifying attributes that further capture the meaning of the statement and characterize the properties of nodes and edges.

Because the basic structure of a KG is generic, the knowledge contained within a KG can be heterogeneous and mutable and still be representable in the graph. The representation of knowledge as simple connections between core entities makes iterative, rapid development of KGs possible. In addition, by leveraging the graph data structure and using various inference strategies, one can infer new edges or connections between nodes in a graph. Ontology-oriented KGs allow deductive inference through logical rules, from basic rules such as the Gene Ontology (GO) 'true path' rule[12] to more sophisticated methods like Description Logic inference[13]. Ontology-oriented KGs are also amenable to machine learning approaches such as embedding in vector space[14], which supports the application of deep neural networks for tasks such as link prediction and node classification. Within the biomedical sciences, ontology-oriented KGs have been used for tasks such as drug repurposing[5], target prioritization[15], and phenotype profile matching [7].

Several ontologies and schemas for representing biomedical knowledge are available. A constellation of domain-specific ontologies from the Open Biological and biomedical Ontology (OBO) Foundry[16] can be used for modeling knowledge. For example, the Semantic Science Integrated Ontology (SIO)[17] is used for representing scientific data and knowledge. The Wikidata Ontology [18] is used by Wikidata for representing knowledge. In terms of schemas, schema.org is used for representing metadata about entities and relationships to other entities. BioSchemas is an extension of schema.org for representing metadata about biological entities.

While existing efforts in modeling knowledge have been valuable, a unified data model that bridges across multiple ontologies, schemas, and data models does not exist. Here, we present Biolink Model as an open-source, universal data model that defines entities and the relationships between these entities within translational science.

# Overview of Biolink Model

Biolink Model is a data model for organizing data in biomedical KGs. The model serves both as a map for bringing together data from different sources under one unified model, and as a bridge between ontological domains.

Biolink Model is composed of several modeling elements, including a hierarchy of defined *Classes*, *Properties* (with defined *Types*), *Predicates*, *Mixins*, and *Associations* (**Table 1**). Domain knowledge in a KG that conforms to Biolink Model is represented using *Associations*. An *Association* minimally includes a subject and an object (Biolink Model classes) related by a Biolink Model predicate, together comprising its core triple (statement or primary assertion). The subject and object of an *Association* are foundational domain concepts (e.g. genes, diseases, chemicals, phenotypes), whose Internationalized Resource Identifiers (IRIs) come from community standard ontologies (e.g. HGNC, MONDO, ChEBI, HPO). The predicate is a Biolink Model element that represents the relationship between the subject and object. *Associations* may also include slots to hold additional metadata about the core triple, primarily information about the provenance and evidence supporting the assertion (**Figure 1**).

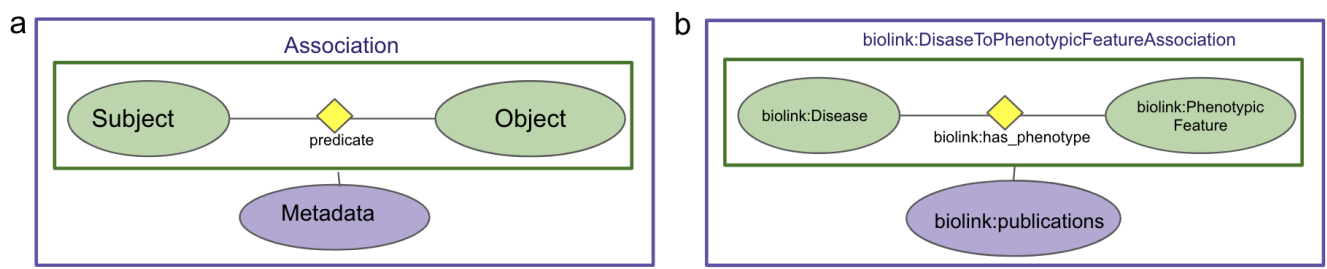

**Figure 1.** An example of an *Association* represented in Biolink Model. In (a), the green ovals represent the subject and object classes, connected by a predicate. Together, the classes and the predicate constitute a statement or '*core triple*' in the model. Edge properties provide further context and qualification to the core triple. The entire diagram, including the core triple and its provenance, represents a Biolink Model '*Association*'. In (b), we see a specific example of a '*biolink:DiseaseToPhenotypicFeatureAssociation*', where the subject is '*biolink:Disease*', the object is '*biolink:PhenotypicFeature*', and the predicate is '*biolink:has_phenotype*'. In addition, the '*biolink:publications*' property (lavender oval) records the provenance of the core triple.

**Table 1**. Biolink Model elements and their definitions.

| Biolink Model Element | Definition | Examples |
|---|---|---|
| Class | High-level types (or categories) representing core biological concepts of interest such as genes, diseases, chemical substances, anatomical structures, and phenotypic features, arranged in a class hierarchy. | biolink:Disease, biolink:PhenotypicFeature, biolink:Gene, biolink:SequenceVariant |
| Predicate | Objects that define the action being carried out by the subject (or named entity) of a core triple and help define how two entities (or classes) can be related to one another. In graph formalism, predicates are relationships that link two instances. Predicates in the Biolink Model all descend from the 'biolink:related_to' predicate. | biolink:has_phenotype, biolink:positively_regulates, biolink:affects, biolink:associated_with, biolink:related_to |
| Node Property | A set of attributes that can be regarded as a characteristic or inherent part of an instance of 'biolink:NamedThing'. | biolink:symbol, biolink:name, biolink:id |
| Edge Property | A set of attributes that can be regarded as a characteristic or inherent part of a statement, association, or edge. | biolink:publications, biolink:has_evidence |
| Core Triple | The domain knowledge of an Association expressed by the subject and object nodes plus the predicate connecting them. | biolink:Disease biolink:has_phenotype biolink:PhenotypicFeature |
| Association | Associations are classes that define a relationship between two domain concepts, constrained and qualified by edge attributes. | biolink:DiseaseToPhenotypicFeatureAssociation, biolink:GeneToDiseaseAssociation |
| Type | A kind of value that tells what operations can be performed on a particular data set. Biolink Model implements common types such as integer and string, but it also defines custom types like quotient and unit. | URI or CURIE, string, integer, biolink:Quotient, biolink:Unit |
| Mixin | Modeling elements used to extend the properties (or slots) of a class, without changing its position in the class hierarchy. Please see the Biolink Model documentation for more information on mixin elements. | biolink:GeneOrGeneProduct, biolink:DiseaseOrPhenotypicFeature |

*Abbreviations: URI = unique resource identifier; CURIE = compact URI.*

Biolink Model aims to address several challenges that obstruct the interoperability between KGs, including: 1) the need for expertise to transform data between tabular, RDF, and graphical models; 2) sparse and/or inconsistent application of ontologies or other controlled vocabularies, as well as differences in the identifiers that are used for storing instances of nodes within a graph; and 3) the lack of a standard approach to model the intersection of ontological domains (e.g., the relationships between genes and diseases).

Using the framework provided by the Linked data Modeling Language (LinkML), Biolink Model is distributed in a variety of formats, including YAML, JSON-Schema, SQL-DDL, Python/Java classes, and RDF. Additionally, Unified Modeling Language (UML) diagrams provide a visual representation of the model. Biolink Model is accessible in frameworks familiar to a wide variety of developers and database engineers. Because the model can be distributed in different formats, the model elements can also be validated using toolchains that already exist (e.g., JSONSchema validation, SQL constraints), thus speeding up the reconciliation of tabular data, ontologies, and graphs.

The biomedical field has been a leader and champion of ontology development. However, this has sometimes led to the development of multiple ontologies or controlled vocabularies for the same domain concept. When this happens, KG creators must identify which vocabulary best suits their needs, as well as understand how to apply concepts from the chosen ontology to their class instances. Biolink Model helps solve this challenge by indicating to users which ontologies should be used for instances of its classes via *identifier prefixes (id_prefixes)*, *mappings,* and *associations*.

Biolink Model describes its classes in a *description* field. Part of the definition of a class is an *id_prefixes* construct. Recognizing that biomedical resources often implement new identifiers for their resource, instead of reusing existing identifiers from other resources[19], Biolink Model encourages reuse of existing ontologies by providing a list of possible ontologies (via *id_prefixes*) in preference order for engineers to use when instantiating model classes. For example, for a disease class, Biolink Model suggests that instances of the class use Mondo (the Mondo Disease Ontology)[20] as the preferred disease vocabulary. The *id_prefixes* modeling construct allows the development of software that can normalize identifiers across data sources. Tools such as the Biomedical Data Translator Node Normalization Service and the Knowledge Graph Exchange (KGX) Framework use the identifier mappings in Biolink Model to return the preferred equivalent identifier when presented with several identifiers that represent the same domain concept but with different namespaces (e.g., NCBIGene versus HGNC gene identifiers).

Each element in Biolink model is mapped, when possible, to equivalent elements in other ontologies or models. Biolink Model employs mapping terms from the Simple Knowledge Organization System (SKOS) namespace to record classes and objects outside the model that can be considered similar in either an *exact, broad, narrow, close,* or *related* manner to the Biolink Model class (e.g., the *broad_mapping* relation implements the skos:*broadMatch*). These mappings render the model and data more computable, allowing software programs to automatically harmonize and connect disparate data sources, thus facilitating interoperability.

Finally, a key feature of Biolink Model is its Association elements. Taking inspiration from successful efforts like Semanticscience Integrated Ontology [17], Biolink Model Association elements establish rules for transforming biomedical knowledge into computable statements and help define how to represent knowledge statements across ontological domains. 'Computable' in this context means that each Biolink Model Association defines the kinds of objects that can participate as a subject or object of a biomedical statement (via domain and range constraints); defines sets of attributes (edge properties described in Table 1 and detailed in the Biolink Model documentation) that are required to properly instantiate a relationship between two domain concepts; and provides a blueprint for registering and maintaining the provenance of each statement. In Web Ontology Language (OWL) [21], Biolink Model Association elements are equivalent to *Axioms*, and in RDF, they are equivalent to *Statements (rdf:Statement)*. Because provenance and evidence are critical components of any data set (and the knowledge represented therein), Biolink Model provides properties capable of tracking evidence and provenance both at the class and association levels.

## Applications of Biolink Model

Biolink Model, while constantly evolving, supports a variety of use cases in clinical, biomedical, and translational science. We highlight several examples here.

### Biomedical Data Translator ('Translator') Consortium

The Translator Consortium has adopted Biolink Model as an open-source upper-level data model that supports semantic harmonization and reasoning across diverse Translator 'knowledge sources'[15]. The model serves a central role in the Translator program and forms the architectural basis of the Translator system, as described below.

The Translator program aims to develop a comprehensive, relational, N-dimensional infrastructure designed to integrate disparate data sources—including objective signs and symptoms of disease, drug effects, chemical and genetic interactions, cell and organ pathology, and other relevant biological entities and relations—and reason over the integrated data to rapidly derive biomedical insights[22]. The ultimate goal of Translator is to augment human reasoning and thereby accelerate translational science and knowledge discovery.

To achieve this ambitious goal, the Translator project assembled a diverse interdisciplinary team and a variety of biomedical data sources, including electronic health record data, clinical trial data, genomic and other -omics data, chemical reaction data, and drug data. However, the Translator data sources were in formats that were not compatible or interoperable. Moreover, groups within the Translator Consortium had integrated the data sources as knowledge sources within independent KGs, but these KGs were developed using different technologies and formalisms such as property graphs in Neo4j and semantically-linked data via RDF and OWL.

In order to interoperate between knowledge sources and reason across KGs, Biolink Model was adopted as the common dialect to provide rich annotation metadata to the nodes and edges in disparate graphs, thus enabling queries over the entire Translator KG ecosystem, despite incompatibilities in the underlying data sources. The result was a federated, harmonized ecosystem that supports advanced reasoning and inference to derive biomedical insights based on user queries.

An example Translator use case involved a collaboration with investigators at the Hugh Kaul Precision Medicine Institute (PMI) at the University of Alabama at Birmingham. PMI investigators posed the following natural-language question to the Translator Consortium: *what chemicals or drugs might be used to treat neurological disorders such as epilepsy that are associated with genomic variants of* RHOBTB2? The investigators noted that *RHOBTB2* variants cause an accumulation of RHOBTB2 protein and that this accumulation is believed to be the cause of the neurological disorder.

To answer the PMI investigator's question, Translator team members structured the following query: *NCBIGene:23221* (CURIE for RHOBTB2) -> *[biolink:entity_regulates_entity, biolink:genetically_interacts_with] -> biolink:Protein, biolink:Gene -> [biolink:related_to] -> biolink:SmallMolecule* (**Figure 2**). Because of the hierarchical structure of the Biolink model, the use of *biolink:related_to* also will return more specific predicates such as *biolink:negatively_regulates* and *biolink:positively_regulates*. The objective was to identify drugs or chemicals that might regulate *RHOBTB2* in some manner and thereby reduce the variant-induced accumulation of RHOBTB2 and associated neurological symptoms. As all nodes and edges within the Translator KG ecosystem are annotated to Biolink Model classes and attributes, a Translator query can be constructed from a natural-language user question and return results across a multitude of independent data sources. In addition, because the model employs hierarchical classes, with inheritance and polymorphism, natural-language queries translated to graph queries using Biolink Model syntax can be constructed at varying levels of granularity and return results from all levels of the hierarchy. Finally, because Biolink Model provides attributes on both edges and nodes that record provenance and evidence for these knowledge statements, each result is annotated with the trail of evidence that supports it.

When Translator team members sent the query to the Translator system, it returned several candidates of interest to PMI investigators, including fostamatinib disodium (CHEMBL.COMPOUND:CHEMBL3989516) and ruxolitinib (CHEMBL.COMPOUND:CHEMBL1789941). A review of the supporting evidence provided by Translator indicates that these are approved drugs that either directly or indirectly reduce or otherwise regulate the expression of *RHOBTB2*. Thus, Biolink Model helped Translator teams bring data together into a single system, thereby reducing the burden on the user to find and manually assemble data from these independent resources.

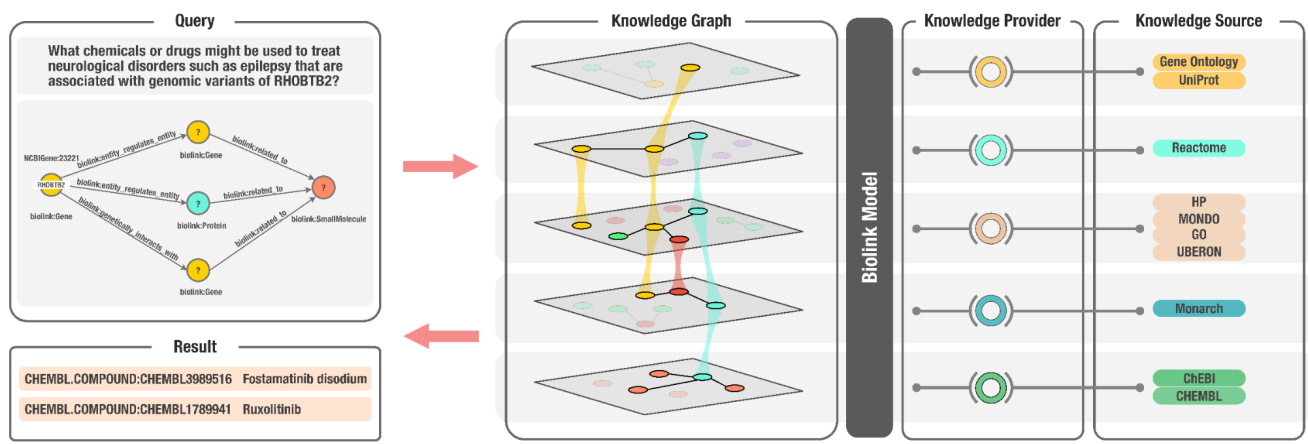

*Figure 2.* An overview of the Translator architecture that supports biomedical KG-based question-answering, including the role of Biolink Model, in the context of an example question. In this example, a user has posed the natural-language question: *what chemicals or drugs might be used to treat neurological disorders such as epilepsy that are associated with genomic variants of RHOBTB2?* The question is translated into a graph query, as shown in the top left panel, which is then translated into a Translator standard machine query (not shown). The KG shown in the second panel from the left is derived from a variety of diverse 'knowledge sources', a subset of which are displayed in the figure, that are exposed by Translator 'knowledge providers'. Biolink Model provides standardization and semantic harmonization across the disparate knowledge sources, thereby allowing them to be integrated into a KG capable of supporting question-answering. In this example, Translator provided two answers or results of interest to the investigative team who posed the question, namely, fostamatinib disodium and ruxolitinib, as shown in the bottom left panel.

## Additional Applications and Reuse of the Biolink Model

While developed in concert with the use cases of the Translator Consortium, Biolink Model has been reused in other applications, including KGX and Knowledge Graph Exchange Archive (KGEA), which rely on universal schemas for data model structure and data integration. In addition, the Illuminating the Druggable Genome (IDG) project uses Biolink Model as a schema for its integrated view of genomic, phenomic, and biochemical data. Similarly, the Monarch Initiative uses Biolink Model as a schema for its integrated view of genomic, phenomic, and biochemical data. Both IDG and Monarch incorporate a broad spectrum of data from a variety of sources, with each source modeling their data using different approaches, independent identifier systems, and heterogeneous data representations. Biolink provides the semantic harmonization required to integrate these disparate data sources. Other initiatives that rely on Biolink Model for data and knowledge harmonization include KG-COVID-19[10], ECO-KG, KG-ENVPOLYREG, and KG-Microbe.

## Discussion

The success of Biolink Model can be attributed to its community—biologists, clinicians, data curators, developers, subject matter experts, and ontologists—all of whom have contributed their requirements, perspectives, and expertise to help build a flexible semantic data model. Biolink Model is under continual development, with frequent releases and a publicly-accessible issue tracker on GitHub. To ensure sustained development of the model, we invite the biomedical community to contribute via GitHub pull requests and use the issue tracker to suggest new features, report problems, or ask questions. (See Supplemental Resources within Supplementary Materials for links to the GitHub repository for Biolink Model, documentation, and other relevant resources.)

Biolink Model provides a blueprint to harmonize existing data sources and accelerate the development of new knowledge by leveraging a multitude of domain and technical expertise, captured in a variety of ontologies and existing models (via semantic mappings), within a single modeling framework that is easy to read, write, reuse and distribute. Moreover, Biolink Model is grounded in semantic web technologies (characterized by classes and slots with their own IRIs, SKOS mappings to existing ontologies, descriptions, identifier prefixes, and domain and range constraints) and captures biomedical expertise as a computable knowledge artifact that can be read and interpreted by both machines and humans alike.

Because Biolink Model is platform-agnostic, open-source, and publicly accessible, and because it can be translated into a variety of data modeling formats, it encourages people from different backgrounds and with different expertise to work together to evolve the model. Most importantly, Biolink Model supports the harmonization of KGs and underlying data sources in a manner that adheres to FAIR principles[23] and facilitates applications across a broad spectrum of biomedical use cases, thereby democratizing and accelerating translational science.

## Conflicts of Interest

All authors declare no conflicts of interest.

## Acknowledgements

The authors are grateful to members of the Publications Committees at the National Center for Advancing Translational Sciences, the National Institute of Environmental Health Sciences, and the National Institute on Aging for their review and approval of the manuscript for publication. Moreover, the authors are appreciative of the unwavering leadership and support provided by the Extramural Leadership Team and the Intramural Research Program at NCATS.

## Funding Support

This work was supported in part by the NCATS Biomedical Data Translator program (other transaction awards OT2TR003434, OT2TR003436, OT2TR003428, OT2TR003448, OT2TR003427, OT2TR003430, OT2TR003433, OT2TR003450, OT2TR003437, OT2TR003443, OT2TR003441, OT2TR003449, OT2TR003445, OT2TR003422, OT2TR003435, OT3TR002026, OT3TR002020, OT3TR002025, OT3TR002019, OT3TR002027, OT2TR002517, OT2TR002514, OT2TR002515, OT2TR002584, OT2TR002520; Contract number 75N95021P00636). Additional funding was provided by the Office of the Director, National Institutes of Health (grant award R24-OD011883), the National Human Genome Research Institute (grant award 7RM1HG010860-02), and the Director, Office of Science, Office of Basic Energy Sciences, of the U.S. Department of Energy under Contract No. DE-AC0205CH11231.

# Supplementary Materials

## Supplemental Resources

**Biolink Model**
Schema, generated objects, and associated code: https://github.com/biolink/biolink-model
Documentation: https://biolink.github.io/biolink-model/
Documentation about edge properties: https://biolink.github.io/biolink-model/docs/edge_properties.html
Documentation about mixins: https://biolink.github.io/biolink-model/guidelines/using-the-modeling-language.html#mixin
Issue tracker: https://github.com/biolink/biolink-model/issues

**Other relevant resources**
SKOS Simple Knowledge Organization System (SKOS): https://www.w3.org/TR/2009/REC-skos-reference-20090818/
Resource Description Framework (RDF): https://www.w3.org/TR/rdf11-concepts/
Neo4J: https://neo4j.com/
schema.org: http://schema.org
Biomedical Data Translator Node Normalization Service: https://github.com/TranslatorSRI/NodeNormalization
Knowledge Graph Exchange Framework: https://github.com/biolink/KGX
Linked data Modeling Language (LinkML): https://github.com/linkml/linkml
Knowledge Graph Exchange (KGX): https://kgx.readthedocs.io/en/latest/
Knowledge Graph Exchange Archive (KGEA): https://github.com/NCATSTranslator/Knowledge_Graph_Exchange_Registry
ECO-KG: https://github.com/Knowledge-Graph-Hub/eco-kg
KG-COVID-19: https://github.com/Knowledge-Graph-Hub/kg-covid-19
KG-ENVPOLYREG: https://github.com/Knowledge-Graph-Hub/kg-envpolyreg
KG-Microbe: https://github.com/Knowledge-Graph-Hub/kg-microbe
KG-OBO: https://github.com/Knowledge-Graph-Hub/kg-obo
KG-IDG (Illuminating the Druggable Genome): https://github.com/Knowledge-Graph-Hub/kg-idg
Monarch Initiative (ingest): https://github.com/monarch-initiative/monarch-ingest, https://github.com/Knowledge-Graph-Hub/sri-reference-kg
Unified Modeling Language (UML): https://www.omg.org/spec/UML/
Ontology Web Language (OWL): https://www.w3.org/OWL/
OBOFoundry: https://obofoundry.org/
SIO: http://www.ontobee.org/ontology/SIO
Wikidata: https://www.wikidata.org/wiki/Wikidata:Main_Page
Bioschemas: https://bioschemas.org/
Yet Another Modeling Language (YAML): https://yaml.org/spec/
JSON-Schema specification: https://json-schema.org/specification.html
Resource Description Framework (RDF): https://www.w3.org/RDF/
Hugh Kahl Precision Medicine Institute (PMI): https://www.uab.edu/medicine/pmi/